# On the Conservation of Number of Nodes and the Consumed Energy in Wireless Sensor Network: A Statistical Mechanics Approach

Chiranjib Patra [a]

[a] *Calcutta Institute of Engineering and Management, Tollygunge,*

*Kolkata-700040, India*

**Abstract**

The dominant dynamics of sensor networks consist of using the energy of the sensor nodes to create the topology of hierarchical clustering using topology control protocols. The topology thus created will always have optimum number of nodes after using a certain amount of energy, vice versa optimum amount of energy expense to use a certain number of nodes. This paper attempts to find a relation between the number of nodes and the energy consumed using statistical mechanics. This relationship thus obtained validates considerably well with the simulation experiments with topology control protocols.

1. **Introduction**

Recent studies in natural sciences has been focusing on an increased interest in the field of complex networks—found in diverse areas such as biological networks ,sensor networks, social networks and other interdisciplinary networks. The application of complex networks as of interacting complex dynamical systems has been often used in explaining the properties of system underlying in the internet and sensor networks. The major challenges lies in understanding the high dimensionality and complex topological structures of these systems.

Conventionally these network problems are investigated by using graph theory [9] but current work shows that the application of Statistical Mechanics facilitate the understanding and modelling of the complex topological characteristics in sensor networks. For example, by mapping nodes of a graph to energy levels and edges as particles occupying that energy level, [1] have shown that complex networks follow Bose statistics and might undergo Bose Einstein condensation despite their irreversible and non-equilibrium nature. In another article [2] have presented a fast community detection algorithm using a q-state Potts model. A detailed review of articles in this area can be found in [3, 4]. Thus the usage of Statistical Mechanics in sensor networks can done in two ways:

1) Analysing the topology of the sensor networks, whose structures lie in between a totally ordered and completely disordered. 2) Developing algorithms, to regulate the local behaviours that through mutual interaction produce desired ensemble characteristics from a global perspective. One of the important issues in the area of sensor networks is the judicious reallocation of available resources such as energy and communication bandwidth.

As the availability of these resources in WSN considerably affect the performance and lifetime of the network, so optimization of the resources becomes necessary[10]. In the present scenario we develop a framework using the concepts of statistical mechanics to relate between the energy consumption and the number of sensor nodes in participation. This relation has been supported using simulation experimental data.

## 2. The Proposed Model for WSN's

We start with the sensor nodes distributed in area under consideration having energy E, but after the invocation of the topology control protocols like A3, CDS- Rule K etc. every sensor nodes expenses a certain amount of energy to develop the connection with other nodes in the vicinity as depicted in figure II. Let the expended energy be e and the connection with each sensor to other sensors with the neighborhood be k. There is also another view for this connections made, these connections are proportional to the energy of the sensor node i.e. more the energy of the node more connections can be made.

So we can consider for a particular expended energy state e1 for a sensor under consideration, there may be k1,k2,k3,......kn connection levels. Thus in the light of statistical physics we can think of these sensors as $N_i$ indistinguishable items which can be distributed among $k_i$ connections, the number of ways in which $N_i$ can be distributed among $k_i$ connections is $\frac{(k_i)^{N_i}}{N_i!}$.

At equilibrium we have at the same energy state a node can exhibit one or more connectivity levels much larger than the nodes found in any one such level. At one time the specifications may be defined as

$N_1$ sensor nodes in the energy state $e_1$ with connectivity level $k_1$
$N_2$ sensor nodes in the energy state $e_2$ with connectivity level $k_2$
..................................................................................................................
..................................................................................................................
$N_i$ sensor nodes in the energy state $e_i$ with connectivity level $k_i$
So the micro state may be defined as the product of the type

$$\Omega = \frac{(k_1)^{N_i}}{N_1!} * \frac{(k_2)^{N_2}}{N_2!} * \ldots \ldots \ldots \frac{(k_i)^{N_i}}{N_i!}$$

At equilibrium using sterling approximation, we evaluate $\Omega$ at maximum we obtain

$$\ln \Omega = \sum N_i \ln \frac{k_i}{N_i} + N$$

Our problem now is to render in $\Omega$ to a maximum subject to the condition that
$\sum N_i = N$.
$\sum N_i e_i = U$.
Using Lagrange multiplier, we have

$$N_i = A k_i e^{-\beta e_i}$$

Where A and β are Lagrange's multiplier

As we have $\sum N_i = N$.
So we can write $\quad N = A \sum k_i \, e^{-\beta e_i}$

Hence
$$\frac{N}{\sum k_i \, e^{-\beta e_i}} = A$$

$$\frac{N}{Z} = A \quad \text{------------------------------------------------ (1)}$$

Where $Z = \sum k_i \, e^{-\beta e_i}$ Where z is called the partition function
Expended energy consideration in sensors

$$E(l, d) = (\alpha_0 + \beta_1 d^2)l$$

$\alpha_0$ = transmitter circuit energy
l= bits
$\beta_1$= transmitter amplifier energy
d= distance of propagation
E(l,d)= energy consumed
Assuming this much energy is required for for a single connection so for $k_i$ connections , so we have
$e_i = k_i * E(l,d)$
Thus we can write,

$$Z = \sum k_i \, e^{-\beta(k_i * E(l,d)_i)}$$

now we can consider replacing the summation sign by integration sign with $dk_i$ limit from zero to infinity (large value of k)

$$Z = \int_0^\infty k_i \, e^{-\beta(k_i * E(l,d)_i)} dk_i$$

$$Z = \frac{1}{(\beta(E(l,d)_i)^2} \quad \text{-------------------------------- (2)}$$

Using equation 1 and 2 we have

$$A = N((E(l,d)_i)^2$$

$$\frac{A}{\beta^2} = NE^2$$

Hence we have  $NE^2 = constant$  ---------------------------- (3)

Now if we apply the above result in hierarchical clustering of sensor networks with the host of topology control protocols like CDS-RuleK, EECDS, A3 to name the few , the above equation 3 may be modified by the principle of conservation

$$N_{Total} E_{Total}^2 = N_{cluster-head} E_{cluster-head}^2 + N_{normal} E_{normal}^2 \quad \text{---- (4)}$$

### 3. Experimental Setup

The experiment was carried using Atarraya [6,7] Java based simulator . Atarraya (Version 1.3) is a simulator for Topology Control algorithms.Copyright (C) 2005-2011  Pedro M. Wightman. In order to find the data set using protocol CDS rule –K. We set upon **deployment options**, **Attaraya**,

**Visualizations**, where we set the following parameters as described in the screen shot. Other tabs or parameters are left as default.

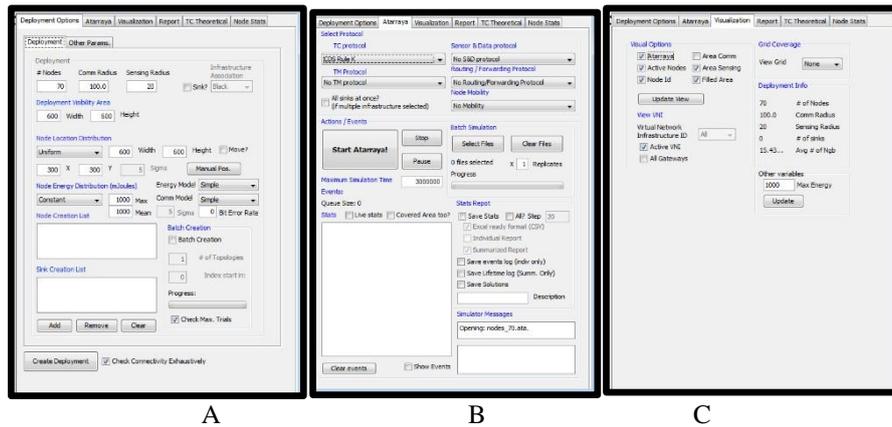

A  B  C

*The screen shots of the Atarraya Simulator for the preparation for the experiment of protocol CDS – Rule K*

**Figure-I**

After the parameters are set as described in the above figure, then we choose the button **StartAtarraya** and we get the display of the **Figure II**. After the total simulation is completed, then we choose the red circles as cluster-heads as defined. Then we go from identifying the cluster-heads with ids' as shown in the figure below, then we count the neighbours, including other cluster-heads for a target id chosen.[6,7]. The various output due to CDS-Rule K [8] protocol are as displayed in the figure below

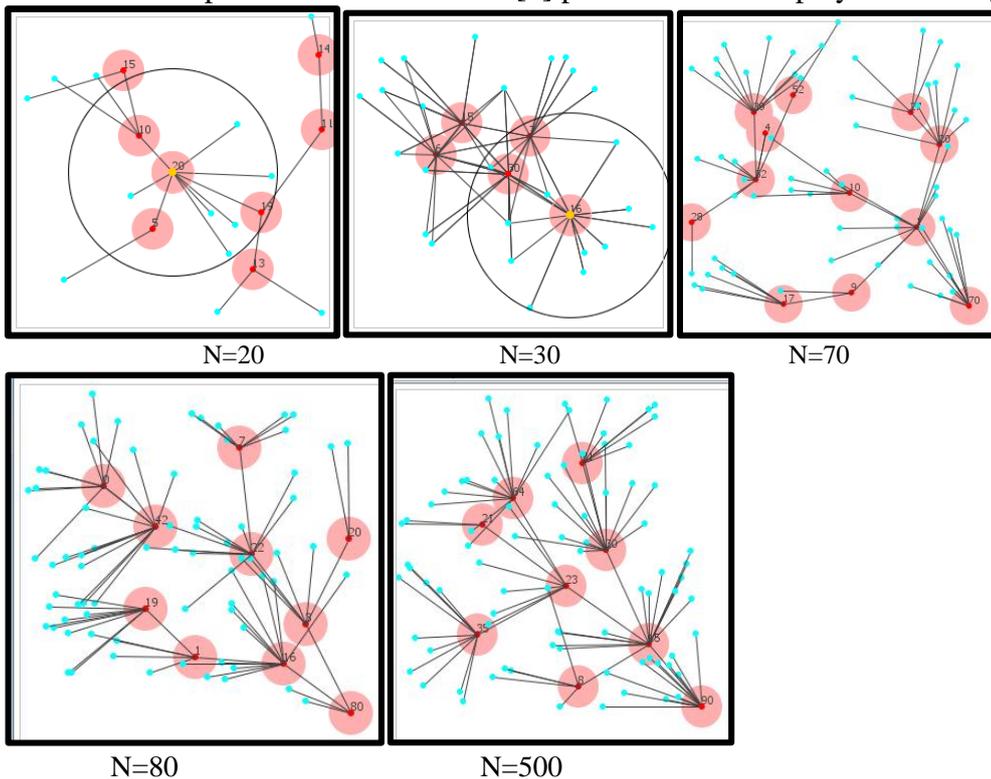

N=20  N=30  N=70

N=80  N=500

*As the output obtained by running the simulator at various capacities I N=100,200,300,400,500 keeping the test area and other parameters fixed*

**Figure-II**

The fixed parameters are as described in the screen shot of Figure-I , except for the number of nodes. The similar experiments for the A3 and EECDS protocols are performed in the simulator and the results are as discussed in the next section.

## 4. Simulation and Results

In order to carry out the relevance of the experimental results we introduce a Q factor which can be easily formulated using equation 4 as

$$Q = \frac{N_{cluster-head}E_{cluster-head}^2 + N_{normal}E_{normal}^2}{N_{Total}E_{Total}^2} \quad \text{------------- (5)}$$

Carrying out the simulation using CDS-RuleK for all collection then combining the theoretical (using equation 15) and experimental values we have the table as described below

| Test Area | $N_{Total}$ | $N_{Cluster-head}$ | $N_{normal}$ | $E_{Total}$ | $E_{cluster-head}$ | $E_{Normal}$ | Q |
|---|---|---|---|---|---|---|---|
| 300X 300 | 20 | 9 | 11 | 0.397177 | 0.327978571 | 0.466375794 | 0.93879004 |
| 300X 300 | 30 | 5 | 25 | 0.323038 | 0.319709 | 0.326367 | 0.986342525 |
| 300X 300 | 40 | 11 | 29 | 0.313583 | 0.283117 | 0.344048 | 0.911682595 |
| 300X 300 | 50 | 13 | 37 | 0.336075 | 0.311768 | 0.360382 | 0.930523279 |
| 300X 300 | 60 | 14 | 36 | 0.366331 | 0.339762 | 0.392899 | 1.12246041 |
| 300X 300 | 70 | 11 | 59 | 5.328688266 | 5.2491713 | 5.408205232 | 0.979731585 |
| 300X 300 | 80 | 12 | 68 | 8.500981322 | 7.882336213 | 9.11962643 | 0.903196574 |
| 300X 300 | 90 | 9 | 81 | 9.920066094 | 9.968660678 | 9.871471511 | 1.007875323 |
| 300X300 | 100 | 11 | 89 | 13.20144444 | 13.3004 | 13.10248889 | 1.011774301 |

**Table-I**

*Dataset generated using CDS-RuleK Protocol*

It can be easily observed that the expected Q value is required to be 1 in order to satisfy equation 4 but however it is well acceptable in the range of ±10% for experimental to theoretical prediction.

Similarly again carrying out the simulation using A3 for all collection, then combining the theoretical (using equation 5) and experimental values we have the table as described below

| Test Area | $N_{Total}$ | $N_{Cluster-head}$ | $N_{normal}$ | $E_{Total}$ | $E_{cluster-head}$ | $E_{Normal}$ | Q |
|---|---|---|---|---|---|---|---|
| 300X 300 | 20 | 6 | 14 | 1.04866 | 1.128012 | 0.969307 | 1.057988 |
| 300X 300 | 30 | 9 | 21 | 1.222795 | 1.300374 | 1.145216 | 1.049021 |

| Test Area | N_Total | N_Cluster-head | N_normal | E_Total | E_cluster-head | E_Normal | Q |
|---|---|---|---|---|---|---|---|
| 300X 300 | 40 | 10 | 30 | 1.402737 | 1.511228 | 1.294246 | 1.076844 |
| 300X 300 | 50 | 9 | 49 | 1.737234 | 1.869262 | 1.605205 | 0.956845 |
| 300X 300 | 60 | 9 | 51 | 1.928685 | 2.17987 | 1.6775 | 1.198135 |
| 300X 300 | 70 | 10 | 60 | 2.085768 | 2.277628 | 1.893907 | 1.140181 |
| 300X 300 | 80 | 9 | 71 | 2.418759 | 2.633707 | 2.203811 | 1.149223 |
| 300X 300 | 90 | 11 | 79 | 9.600267 | 9.229915 | 9.970619 | 0.94359 |
| 300X300 | 100 | 13 | 87 | 2.853738 | 2.882799 | 2.824677 | 1.015195 |

**Table-II**

*Dataset generated using A3 protocol*

It can again be easily observed that the expected Q value required is 1, in order to satisfy equation 4, but however it is well acceptable in the range of ±14% for experimental to theoretical prediction

Similarly again carrying out the simulation using EECDS topology control algorithm for all collection, then combining the theoretical (using equation 5) and experimental values we have the table as described below

| Test Area | N_Total | N_Cluster-head | N_normal | E_Total | E_cluster-head | E_Normal | Q |
|---|---|---|---|---|---|---|---|
| 300X 300 | 20 | 4 | 16 | 1.338583 | 1.156991 | 1.520175 | 0.8466 |
| 300X 300 | 30 | 7 | 23 | 4.90182 | 5.018501 | 4.785139 | 1.025456 |
| 300X 300 | 40 | 9 | 31 | 3.876517 | 3.510025 | 4.243009 | 0.898526 |
| 300X 300 | 50 | 8 | 42 | 6.304055 | 6.112637 | 6.495473 | 0.959493 |
| 300X 300 | 60 | 8 | 52 | 7.888325 | 7.93423 | 7.84242 | 1.008574 |
| 300X 300 | 70 | 10 | 60 | 7.485043 | 7.326598 | 7.643487 | 0.970225 |
| 300X 300 | 80 | 8 | 72 | 10.59175 | 10.20065 | 10.98284 | 0.943002 |
| 300X 300 | 90 | 10 | 80 | 13.7097 | 12.69144 | 14.72796 | 0.892019 |
| 300X300 | 100 | 10 | 90 | 17.30591 | 15.9817 | 18.63012 | 0.886302 |

**Table-III**

*Dataset generated using EECDS protocol*

It can again be easily observed that the expected Q value required is 1, in order to satisfy equation 4, but however it is well acceptable in the range of ±12% for experimental to theoretical prediction

## 5. Results and Discussions

The relationship discussed in equation 4 seems working well in a fixed test condition (the area of the test space, the density of the nodes, the energy distribution, and the node distribution.) i.e. the test condition of LHS and RHS of equation 4 should be same. This equation can be used to predict the number of cluster nodes and the normal nodes after the expense of a certain energy. This prediction will in turn allow the network engineers to have a first information of any intrusion or an energy hole in network.

Another important point about the equation 3 , if modified to find amongst the total energy values to the nodes from Table I or Table II or Table III (like $N_{20}E^2_{20} = N_{60}E^2_{60}$) ,then this will not be valid because this requires the ensemble to be same when the network performing with 20 nodes or 60 nodes .This directly indicates the constant from equation 3 is different for different simulations situation.

## 6. References

1. G. Bianconi and A.-L. Baraba´si, ''Bose-einstein condensation in complex networks,'' Phys. Rev.Lett., vol. 86, no. 24, pp. 5632–5635, June 2001.

2. J. Reichardt and S. Bornholdt, ''Detecting fuzzy community structures in complex networks with a Potts model,'' Phys. Rev. Lett., vol. 93, no. 21, p. 218701, Nov 2004.

3. R. Albert and A.-L. Baraba´si, ''Statistical mechanics of complex networks,'' Rev. Mod. Phys.,vol. 74, no. 1, p. 47, 2002

4. S. Strogatz, ''Exploring complex netwoks,'' Nature(London), vol. 410, pp. 268–276, March 2001

5. Pedro Wightman and Miguel A. Labrador, "A3: A Topology Control Algorithm for Wireless Sensor Networks," IEEE Globecom 2008, November 2008.

6. Pedro Wightman and Miguel A. Labrador, "Atarraya: A Simulation Tool to Teach and Research Topology Control Algorithms for Wireless Sensor Networks", ICST 2nd International Conference on Simulation Tools and Techniques, SIMUTools 2009, February 2009.

7. Topology Control in Wireless Sensor Networks. Tutorial given in IEEE Latincom 2010

8. Miguel A. Labrador and Pedro Wightman "Topology Control in Wireless Sensor Networks - with a companion simulation tool for teaching and research ", Springer Science + Business Media B.V. 2009. ISBN: 978-1-4020-9584-9.

9. D. B. West, Introduction to Graph Theory, 2nd ed. Prentice Hall, 2000.

10. I.F. Akyildiz, W. Su*, Y. Sankarasubramaniam, E. Cayirci 'Wireless sensor networks: a survey' Computer Networks 38 (2002) 393–422